# Impurity-level induced broadband photoelectric response in wide-band semiconductor $SrSnO_3$

**Author:**

Yuyang Zhang[12], Lisheng Wang[13], Weijie Wu[24], Zhaoyang Wang[123], Fei Sun[123], He Jiang[123], Bangmin Zhang[123]✉ & Yue Zheng[123]✉

**Affiliations:**

[1]School of Physics, Sun Yat-sen University, Guangzhou 510275, China;

[2]Centre for Physical Mechanics and Biophysics, school of Physics, Sun Yat-sen University, Guangzhou 510275, China;

[3]State Key Laboratory of Optoelectronic Materials and Technologies, School of Physics, Sun Yat-sen University, Guangzhou 510275, China;

[4]School of Systems Science and Engineering, Sun Yat-sen University, Guangzhou 510275, China;

**Abstract**

Broadband spectrum detectors exhibit great promise in fields such as multispectral imaging and optical communications. Despite significant progress, challenges like materials instability, complex manufacturing process and high costs still hinder further application. Here we present a method that achieves broadband spectral detect by impurity-level in $SrSnO_3$. We report over 200 mA/W photo-responsivity at 275 nm (ultraviolet C solar-bind) and 367 nm (ultraviolet A) and ~ 1 mA/W photo-responsivity at 532 nm and 700 nm (visible) with a voltage bias of 5V. Further transport and photoluminescence results indicate that the broadband response comes from the impurity levels and mutual interactions. Additionally, the photodetector demonstrates excellent robustness and stability under repeated tests and prolonged exposure in air. These findings show the potential of SSO photodetectors and propose a method to achieve broadband spectrum detection, creating new possibility for the development of single-phase, low-cost, simple structure and high-efficiency photodetectors.

## Introduction

Due to high melting points, high breakdown electric fields and other properties of wide-bandgap semiconductors ($\boldsymbol{E}_g$ > 2.3 eV), they are generally used in high-frequency and high-power devices[1–3]. Additionally, the absorption edge in the ultraviolet (**UV**) wavelength range makes wide-bandgap semiconductors having potential for **UV** photodetectors. Usually, there are two types of photodetectors: One is special wavelength photodetector, which requires high sensitivity and single wavelength response used for environment monitoring, astronomy and industrial process monitoring; The other is broadband photodetector, which can detect a wide range light from ultraviolet (**UV**) to visible and even near infrared (**NIR**), and could be used in multispectral imaging, optical communications, surveillance and night vision equipment in our daily lives. Although photodetectors based on single-phase semiconductors normally detect specific region in the **UV**-visible-**NIR** range due to the light absorption limitation, much efforts focu on the broad spectral response with multilayer[4–6] and vertical/lateral heterojunctions[7,8]. However, these complicated structures hinder them from application, and broadband photodetectors using single-phase semiconductors could be beneficial for practical application.

Some different materials and methods had come up for broadband photodetectors[9–13]. Gapless band structure of graphene offers the capability of ultrawide band detection[14,15]; 2D materials transition metal dichalcogenides (TMDs), such as $MoS_2$, $MoSe_2$ and $WSe_2$ are reported due to the large light absorption and ultrathin thickness[4,6]; Directed pyroelectric effect termed photo-pyroelectric effect could be excited through light irradiation, which is also used for broadband photodetectors[12]. However, these materials often face challenges such as environmental instability, complex production methods, and high costs. In principle, the band edge absorption in semiconductor mainly contribute to the special wavelength photodetectors with high responsivity; then the impurity energy level for doped wide-bandgap semiconductors also could attend the photodetection process[11,16–18]. Especially, the defect-defect interactions could induce novel in-gap electronic structures[19], which might be useful for broadband photodetectors.

$SrSnO_3$ (SSO) in perovskite structure normally acts as the transparent electrode, which is also a promising wide-bandgap semiconductors due to its rich phase structures, large absorption coefficient and high charge carrier mobility[20–23]. Due to the close relationship between the electronic transport mechanism with the device performance, understanding the transport process is essential. Some different mechanism has been reported to study its high mobility in bulk SSO and SSO thin film[24–27], including 2D weak localization (2D WL), 3D electron-electron interaction (3D EEI) and electron-electron scattering[28–30]; In addition, phase transition of SSO film has also been reported by Wang, and higher mobility after phase transition under compressive strain was observed in the range of 180 K – 260 K[24]; Further work reveals the complex correlation between the impurity level and the defects[19]. Exploring the novel phase of SSO and corresponding application in broadband photodetectors might extend the application of SSO in perovskite structure.

In this work, detailed transport mechanism at different temperature was investigated and the complex intra-band impurity absorption was explored in broadband photodetectors. We discovered phase transitions in La-doped $SrSnO_3$ (LSSO) films grown on high compressive strain substrate $SrTiO_3$ (STO) at 88 K and 180 K, and the new phase transition at 88 K significantly influences the conductive mechanism and transport properties of the LSSO film with several-fold increase in mobility. Further photoluminescence (PL) study reveals that the deep-level states were responsible for the phase transition and broad spectral response of LSSO-based metal-semiconductor-metal (MSM) photodetectors. This work demonstrated the LSSO is a good candidate for broadband photodetector and provided a simple way to achieve the photodetector.

**Method**

Target of $La_{0.03}Sr_{0.97}SnO_3$ (LSSO) was prepared by conventional high-temperature solid-phase reaction using $La_2O_3$, $SrCO_3$, $SnO_2$ as raw materials. These powders were mixed and then sintered at 1100 °C for 12 h in air to decompose carbonates. After sintering the powder was again ground, and pressed into 1-inch target, and calcined at a temperature of 1200 °C for 12 h. (001) $SrTiO_3$ (STO) substrates were cleaned by deionized water and acetone and then stuck on the heater with silver glue. Pulsed laser deposition (PLD) approach was employed to grow LSSO films on (001) STO substrates at 750 °C, 770 °C, and 790 °C under oxygen pressure of 0.2 Pa. We used KrF exciter laser (248 nm) with a laser energy density of 0.6 $J/cm^2$ and set the pulse repetition rate at 1 Hz, 1000 pulse. After deposition, the films were annealed under an oxygen pressure of $10^{-5}$ Pa at the growth temperature for 20 minutes and finally cooled to room temperature at 15 °C/min.

X-ray diffractometer (XRD, PANalytical X'Pert PRO) was employed to obtain $\theta$-$2\theta$ line scans of all samples. The crystallographic texture of the films was studied using a four-circle diffractometer at the Singapore Synchrotron Light Source (SSLS), Singapore. Thickness of films were calculated by scanning electron microscope (SEM, Quanta 250 FEG). The surface of films was measured by atom force microscope (AFM, MPF-3D infinity). Resistance-Temperature, Hall effect and magnetoresistance measurements were all completed by physical property measurement system (PPMS, DynaCool 9). Photoluminescence of films were checked by steady-state and transient fluorescence spectrometer (FLS, FLS1000) at Sun Yat-Sen University instrumental analysis & research center. Photo-dark current was measured by Keithley 4200 semiconductor characterization system, while the laser (375 nm) was excited by multi-wavelength laser (DLA3510). Photodetectors test was characterized by Keysight 2902 A, using Xe lamp (150 W) as the light source.

**Results & Discussion**

The film thickness is 22 nm from the deposition rate in Fig. S1 of supplementary information (SI), and Fig. 1(a) shows $\boldsymbol{\theta}$-2$\boldsymbol{\theta}$ X-ray diffraction scans for LSSO films grown and annealed at 750, 770 and 790 °C on (001) STO substrate, which shows the 00*l* preferred-orientation. The bulk lattice constant of SSO is 4.03 Å, and the films experience in-plane compressive strain; The averaged out-of-plane lattice constant of films increases with increasing temperature, from 4.08 Å to 4.10 Å. The reciprocal space maps (RSMs) were utilized to acquire more information in Fig. 1(c)-1(d). The calculated in-plane and out-plane lattice parameters of the LSSO film is $\boldsymbol{a} = \boldsymbol{b} = 3.989$

Å and $c$ = 4.096 Å for film grown at 770 °C. The pseudo-cubic (002) and $(\bar{1}03)$ RSMs illustrate the existence of strain relaxation in LSSO film, which is also been shown in the asymmetrical (002) peak of $\theta$-$2\theta$ curves. Besides the lattice constant, the half-integer diffraction at room temperature was shown to measure the $SnO_6$ oxygen octahedral rotation in Fig. 1(e). According to Glazer's notation[31], the $SnO_6$ rotation pattern is $a^-a^-c^+$, which is consistent with Wang's work.[24] The roughness of film surface is ~ 0.35 nm in Fig. 1(b), which is favorable for device fabrication.

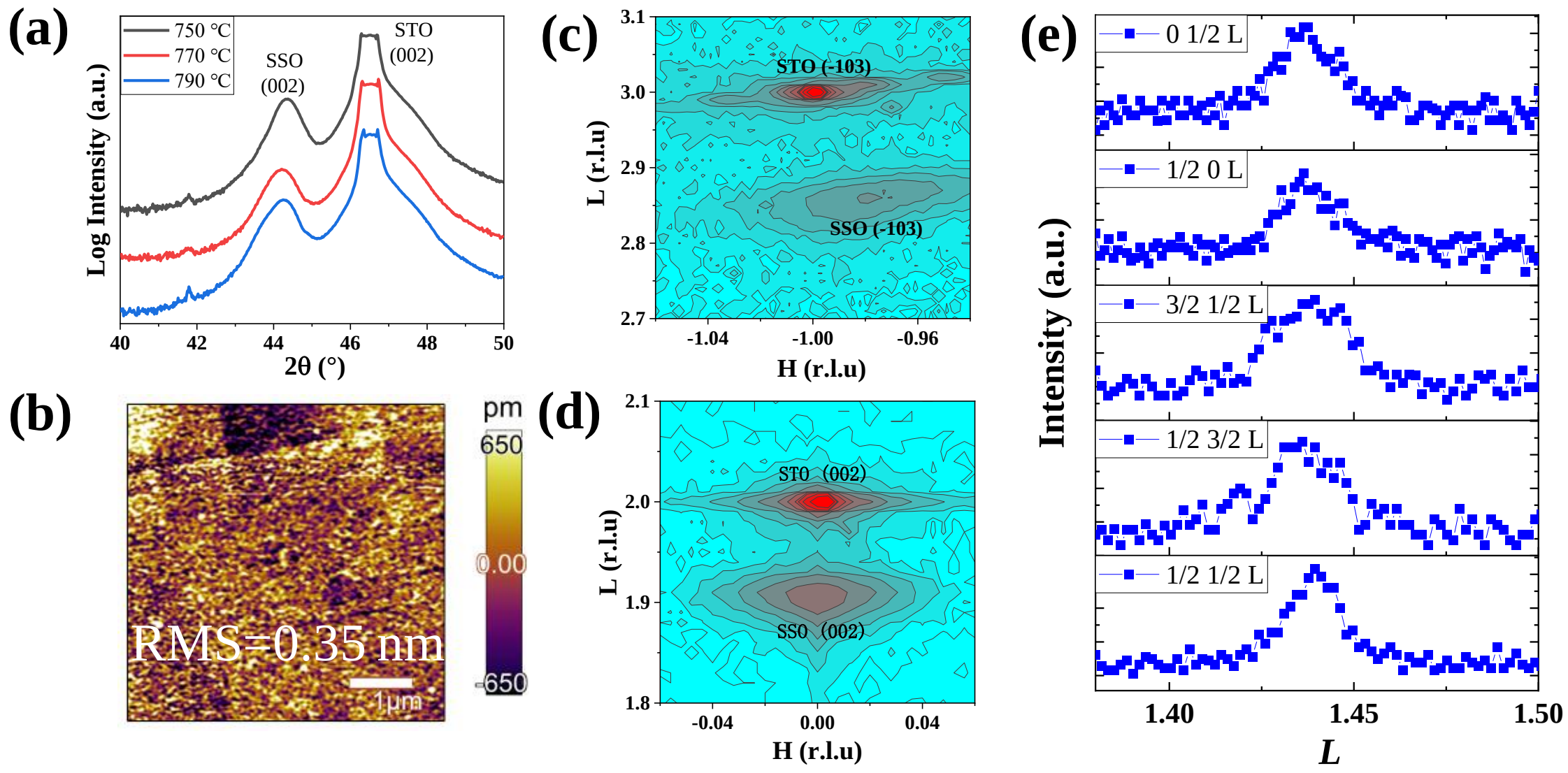

**Fig. 1 configure of LSSO films grown on (001) STO. a** $\theta$-$2\theta$ X-ray diffraction (XRD) of LSSO films. **b** Surface of 750 °C sample measured by AFM. Reciprocal space mapping (RSM) around (-103) **c**, (002) **d**, and **e** half-integer diffraction of 770 °C sample. All the tests are measured at room temperature.

The transport properties of LSSO films are essential parameters for device application, and the resistivity-temperature ($\rho$-$T$) curves is shown in Fig. 2(a) for three films. There are two phase transitions around 88 K ($T_1$) and 180 K ($T_2$) for all films, which is obvious from the derivative curves of $\rho$-$T$ in Fig. S2. Similar phase transition around 180 K has been reported, which is due to the compressive strain.[24] However, the phase transition at $T_1$ ~ 88 K in this work is previously unrecorded in our knowledge and exhibits unique properties. The temperature-dependent carrier density ($n$) and mobility ($\boldsymbol{\mu}$) are shown in Fig. 2(b)-2(c), respectively, with the original Hall test data presented in Fig. S2 of SI. Both $\boldsymbol{n}$ and $\boldsymbol{\mu}$ remain nearly constant across $\boldsymbol{T}_2$, consistent with previous report[24,32,33]. However, film grown at 750 °C exhibits a significant decline in carrier density $\boldsymbol{n}$ below $\boldsymbol{T}_1$, alongside a notable increase in mobility $\boldsymbol{\mu}$. As the temperature falls below $\boldsymbol{T}_1$, carrier density $\boldsymbol{n}$ diminishes, reaching its minimum around 25 K, while $\boldsymbol{\mu}$ exhibits an opposite trend. This trend reverses as the temperature decreases further, with $\boldsymbol{n}$ beginning to increase and $\boldsymbol{\mu}$ to decrease. A similar trend is observed for the film grown at 770 °C. Upon reaching 25 K, a significant decline in $\boldsymbol{n}$, up to 80% and 53% for films grown at 750 °C and 770 °C sample, respectively, was shown compared to that at room temperature; simultaneously, $\boldsymbol{\mu}$ increases by 565% and 133%, respectively. While for film grown at 790 °C, the $\boldsymbol{n}$ and $\boldsymbol{\mu}$ keeps almost constant with fluctuation in the whole temperature range, without obvious changes in $\boldsymbol{n}$ and $\boldsymbol{\mu}$ below $\boldsymbol{T}_1$.

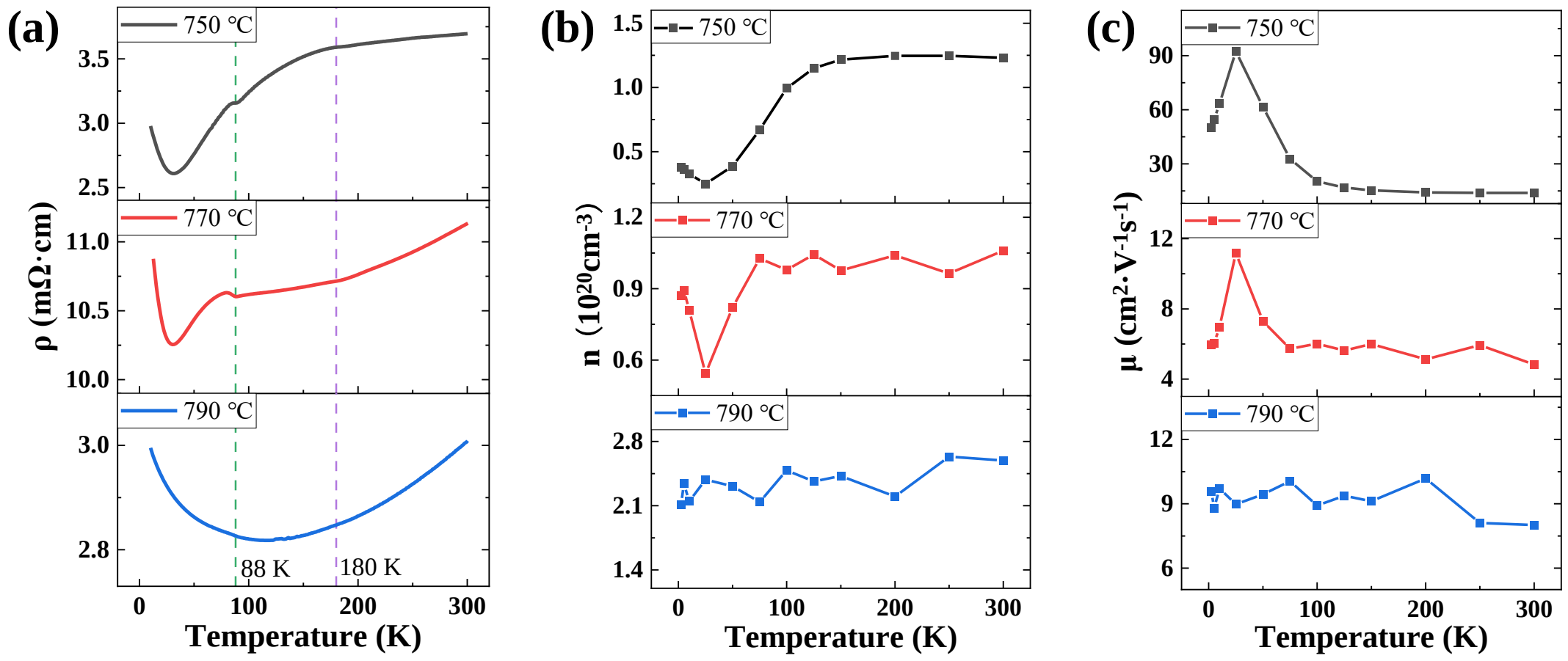

**Fig. 2 Transport properties of LSSO films. a** resistivity-temperature ($\boldsymbol{\rho}$-$\boldsymbol{T}$) curve, **b** carrier density-temperature ($\boldsymbol{n}$-$\boldsymbol{T}$) curve, and mobility-temperature ($\boldsymbol{\mu}$-$\boldsymbol{T}$) curve of three LSSO films.

Then the $\boldsymbol{\rho}$-$\boldsymbol{T}$ and magnetoresistance (MR) ($\frac{R(B)-R(0T)}{R(0T)} \times 100\%$) curves of these three samples were analyzed in Fig. 3(a)-3(f). Red solid lines represent $\boldsymbol{\rho \sim lnT}$ fitting curve at low temperature (below 25 K in films grown at 750 °C and 770 °C, below 100 K in film grown at 790 °C) in Fig 3(a)-3(c), and there is obvious negative MR at low temperature in Fig. 3(d)-3(f), which can be caused by quantum corrections effect-2D weak localization (WL)[28,29,34]. In addition, we measured the temperature-dependent MR ($\frac{R(3T)-R(0T)}{R(0T)} \times 100\%$) of three samples in Fig. 3(g), which also shows negative MR at low temperature range and a bulge around $\boldsymbol{T}_1$. Magnetic field will weaken the influence of 2D WL[35–37], and caused a negative MR. The ***ρ-T*** curves under magnetic field of three samples are shown in Fig. S3.

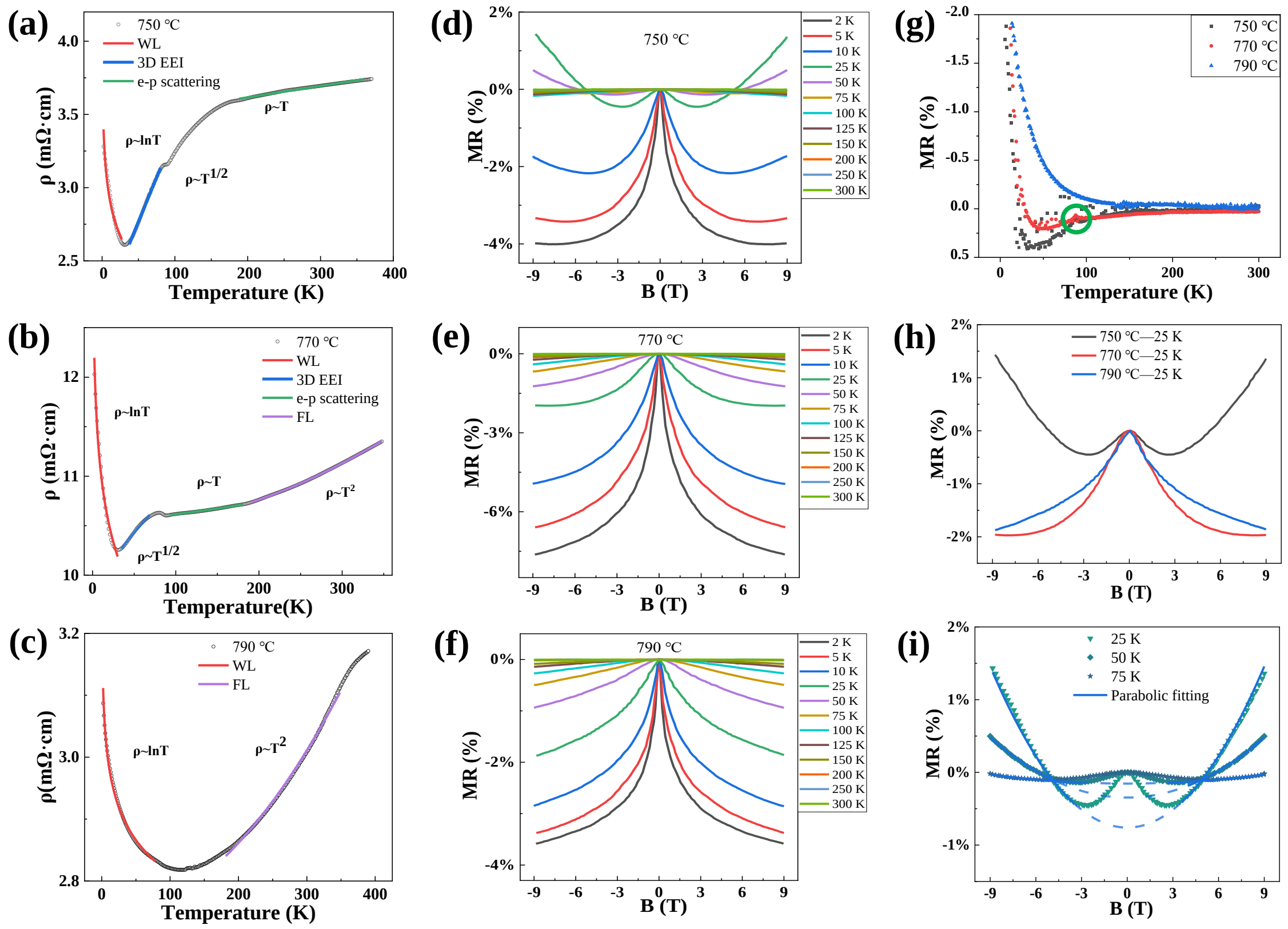

**Fig. 3 magnetoresistance and different mechanism of SSO films.** Different mechanism and corresponding fitting in different temperature interval of 750 °C sample **a**, 770 °C sample **b** and 790 °C sample **c**. Magnetoresistance-magnet field curve at different temperature of 750 °C sample **d**, 770 °C sample **e** and 790 °C sample **f**. **g** MR ((R(3T)-R(0T))/R(0T) × 100%)-temperature curve of three LSSO films. **h** Abstracted MR-temperature curve under 25 K of three samples. **i** Abstracted MR-magnet field curve under 25, 50 and 75 K and parabolic fitting caused by 3D EEI.

With the increase in temperature, the transport mechanism changes. In Fig. 3(a)-

3(b), the blue solid lines are fitting curves of $\boldsymbol{\rho \sim T^{1/2}}$, corresponding to the 3D electron-electron interaction (EEI) model around 25 K < $\boldsymbol{T}$ < 88 K[34,38], which is likely due to the intensity of disorder[39,40]. When the carrier concentration $\boldsymbol{n}$ increases, the screening factor in 3D EEI tends to increase. This trend mainly arises because the increase in $\boldsymbol{n}$ enhances the screening effect among electrons, thereby weakening the direct Coulomb interaction between eletrons[41–45]. Due to the larger drop in $\boldsymbol{n}$ below $\boldsymbol{T}_1$, the enhancement of 3D EEI is more pronounced in the film grown at 750 °C than in the film grown at 770 °C[42,46]. With decreasing temperature, the MR changes gradually from negative MR to positive MR as shown in Fig. 3(h) for film grown at 750 °C, which could be well fitted by $\boldsymbol{\rho \sim H^2}$ caused by 3D EEI[42,47–49]. But negative MR caused by 2D WL also domains under low magnet field at 25 K in Fig. 3(h)-3(i). Moreover, the positive MR in the film grown at 750 °C is stronger than in the film grown at 770 °C, as shown in Fig. 3(g). Conversely, the film grown at 790 °C, which has a higher $\boldsymbol{n}$, does not exhibit positive MR, indicating suppression of the 3D EEI effect.

The significant changes in $\boldsymbol{n}$ and $\boldsymbol{\mu}$ around $\boldsymbol{T}_1$ ~ 88 K, coincide with the temperature where 3D EEI dominate for films grown at 750 °C and 770 °C; and $\boldsymbol{n}$ reaches its minimum at 25 K, where the intensity of 3D EEI is at its peak, leading to the transformation from negative MR to positive MR. However, in films grown at 750 °C and 770 °C, 2D WL becomes dominant below 25 K with increasing $\boldsymbol{n}$, resulting in prevailing negative MR from 25 to 2 K. In contrast, for the sample grown at 790 °C, a broad temperature range up to ~ 100 K is observed where weak localization remains dominant. In this range, 3D EEI effects are not apparent, and $\boldsymbol{n}$ does not undergo significant changes (further discussion follows).

In the film grown at 750 °C, the resistivity above 180 K could be fitted by $\boldsymbol{\rho} \sim \boldsymbol{T}$ (green line), revealing the dominating electron-phonon scattering; The range between 88 K and 180 K, corresponds to the transition range between 3D EEI and electron-phono scattering. In the film grown at 770 °C, the electron-phonon scattering dominates between 100 K < $\boldsymbol{T}$ < 180 K, and above 180 K the $\boldsymbol{\rho} \sim \boldsymbol{T}^2$ fitting curve (purple line) corresponds to the Fermi-liquid (FL) behavior, indicating electron-electron scattering as domain scattering mechanism[28,50]. In the film grown at 790 °C, above 100 K, only the FL behavior appears with dominating electron-electron scattering. For these samples grown at different temperatures, the loss of element Sn might be different, which might affect the lattice constant as revealed by XRD in Fig. 1. The different ratio of Sr/Sn is demonstrated to influence carrier density $\boldsymbol{n}$ by Wang[28], and correlated materials properties[51]. With the diversity of the transport properties, the electronic structure is investigated below.

The steady-state photoluminescence (PL) spectra of films grown at 750 °C and 790 °C over a wide temperature range were measured in Fig. 4 and S4. The temperature-dependent PL spectrum at excitation wavelength $\lambda_{ex} = 275\ nm$ (~ 4.51 eV) in Fig. S4 shows peak shift around ~ 180 K, corresponding to the phase transition at $\boldsymbol{T}_2$; while the temperature-dependent PL centering at 580 nm with excitation wavelength $\lambda_{ex} = 367\ nm$ (~ 3.38 eV) in Fig. 4(a)-4(b) show a sudden drop ~ 88K, corresponding to the phase transition at $\boldsymbol{T}_1$. To get further information, we tested the excitation spectrum of film grown at 750 °C at emission wavelength $\lambda_{em} = 580\ nm$ in Fig. 4(c). The detailed emission spectra at $\lambda_{ex} = 275\ nm$ and $\lambda_{ex} = 367\ nm$ were shown in Fig. 4(d)-4(e). The absorption 1&2 (A1&2) around ~ 275 nm in Fig. 4(c) corresponds to the inter-band excitation, from which the emission PL spectrum is shown in Fig. 4(d) as emission 1&2 (E1&2). An additional absorption 3 (A3) at 367 nm excitation increases gradually below 88 K in Fig. 4(c), corresponding to Emission 3 (E3) in Fig. 4(e). Combining above transport properties, these results suggest the phase transition below 88 K modify the electronic structure with enhanced photon excitation. Additionally, we found some wide and consecutive absorption peaks in Fig. 4(c), which can be used for impurity excitation and broadband spectrum detection.

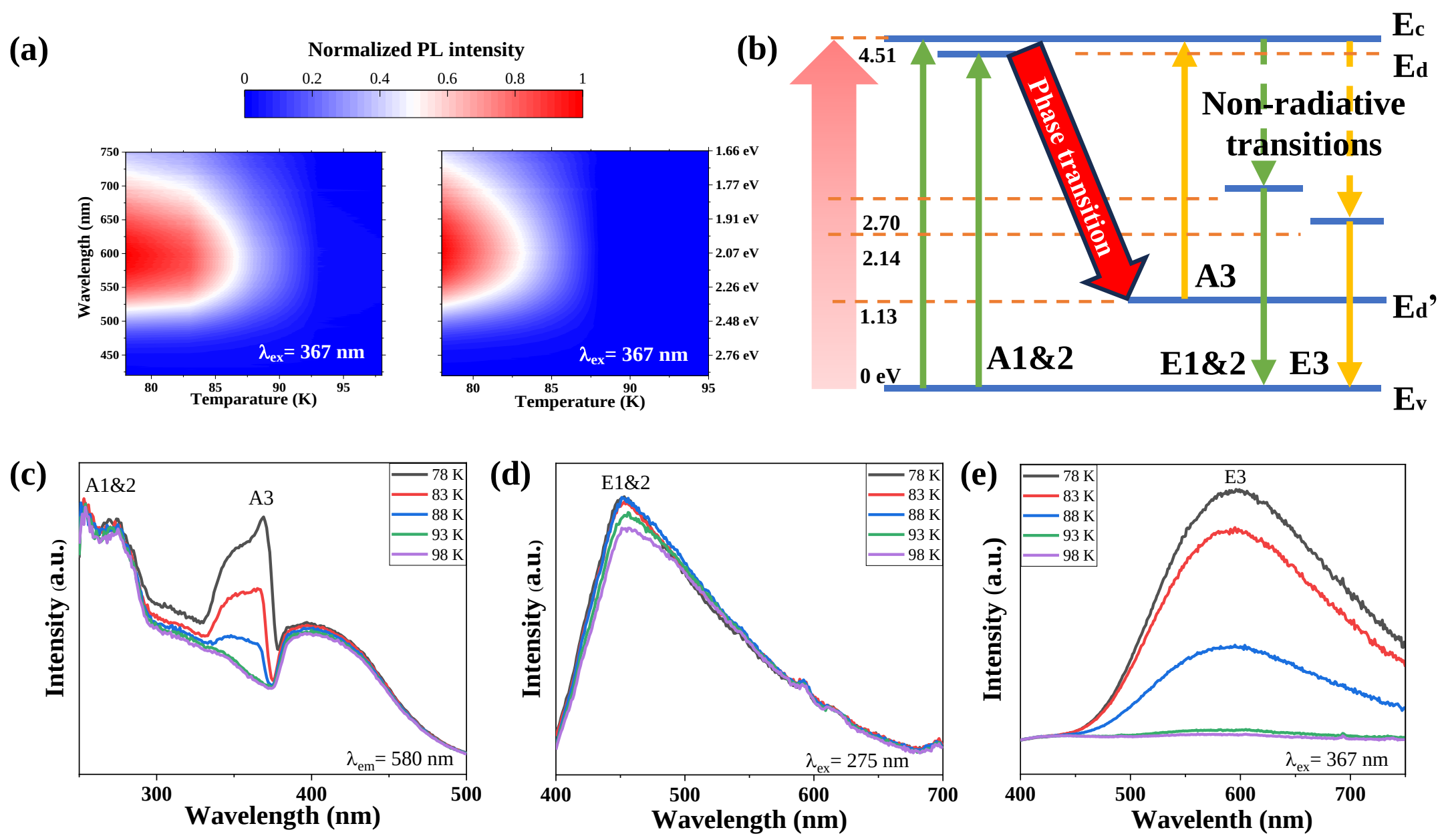

**Fig. 4 Photoluminescence (PL) spectra and energy level model. a** Normalized PL intensity graph under 367 nm excitation from 78 K to 98 K of 750 °C and 790 °C sample. **b** Energy level model though PL tests. **c** Absorption spectrum from 78 K to 98 K at 580 nm emission of 750 °C sample. **d** Emission spectrum from 78 K to 98 K under 275 nm excitation of 750 °C sample. **e** Emission spectrum from 78 K to 98 K under 367 nm excitation of 750 °C sample.

Based on above transport and PL properties, a model was proposed in Fig. 4(b). At high temperature $\boldsymbol{T} > \boldsymbol{T}_1$, the carrier comes from donner level, shown as $\boldsymbol{E}_d$ in Fig. 4(b). The $\boldsymbol{E}_d$ is shallow donor-level, and the $\boldsymbol{n}$ remains stable from 100 K to 300 K in Fig.

2(b). However, the PL intensity at 580 nm under 367 nm excitation increases markedly below $\boldsymbol{T}_1$ as shown in Fig. 4(c) and Fig. 4(e). Below $\boldsymbol{T}_1$, the effect of in-gap deep-energy level $\boldsymbol{E}_d$' increases significantly, which is closed to valence band and might due to defect-defect interactions of oxygen vacancy and La ions[19]. Then, electrons in valence band might jump to $\boldsymbol{E}_d$', and create holes in the valence band, which could recombine with electrons and induce the drop of $\boldsymbol{n}$ in films grown at 750 °C and 770 °C below $\boldsymbol{T}_1$. For film grown at 790 °C with relatively higher carrier density compared to other two films, the fermi level might be high enough and the effect of $\boldsymbol{E}_d$' is suppressed as reported in Ref. 52.

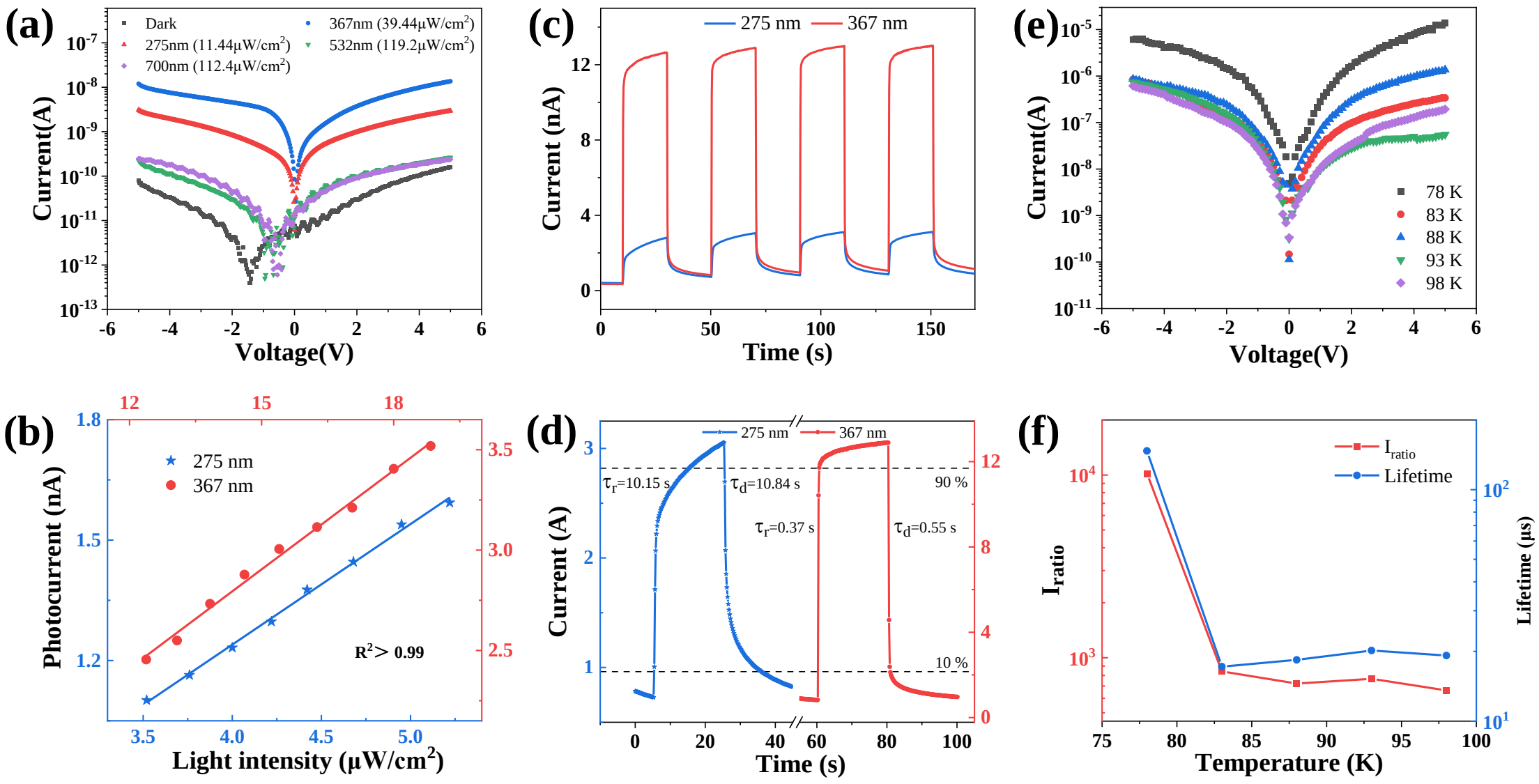

**Fig. 5 Performance tests of LSSO metal-semiconductor-metal (MSM) photodetector device.**
**a** Current-voltage characteristic of LSSO photodetector under 275, 367, 532 and 700 nm light (different photo powder limited by source) with corresponding dark current for comparison at room temperature. **b** Photocurrent under 5V bias as a function of light intensity for SSO photodetectors. **c** Current–time characteristics of LSSO photodetectors under ON/OFF modulated illumination measured at 5 V for the dynamic response and reproductivity evaluation. The monochromatic illumination source has the wavelength of 275 nm (367 nm) and an intensity of 11.44 μW/cm$^2$ (39.44 μW/cm$^2$). **d** Temporal response of the photodetector under 275 and 367 nm excitation. **e** Current-voltage characteristics of LSSO photodetector under 375 nm excitation (203 mW power from laser device) from 78 K to 98 K. **f** Ratio of photocurrent ($\boldsymbol{I}_P$) / dark current ($\boldsymbol{I}_d$) and the lifetime of carrier from 78 K to 98 K.

With the multiple energy levels in doped SSO, then the metal-semiconductor-metal

(MSM) photoelectric detector by LSSO film has been fabricated. Metal Pt (50 nm) is grown on LSSO film to form Schottky contact. The performance of the photodetector at room temperature is shown in Fig. 5(a)-5(d). Fig. 5(a) shows the dark and photo-current under 275, 367, 532 and 700 nm wavelength light excitation, separately, and the ratio of photocurrent/dark current exceeds two orders of magnitude even in a weak light intensity (36 $\mu W/cm^2$) at 367 nm. The responsivity of photodetector under 532 nm and 700 nm is weaker than 275 and 367 nm as shown in Fig. S6[9,13,12], so mainly we discuss the performance of 275 and 367 nm excitation here. Fig. 5(b) shows a linear correlation between the photocurrent of the device and the excitation light intensity, which is attributed to the increased absorption of photons and the generation of photo-excited carriers with rising light intensity. The device exhibits a high degree of linearity with an $R^2 > 0.99$ across the entire range of light intensities, suggesting less of light-induced degradation in the device. Due to the limitation of source (xenon lamp), the light intensity at 367 nm is about ~ 4 times of that at 275 nm, and the photocurrent under 367 nm is also about ~ 4 times of that at 275 nm. With the above linear correlation between the photocurrent and the excitation light intensity, it is anticipated under the same light intensity, the photocurrent under 275 nm should be similar as that under 367 nm in Fig. 5(a). The time-dependent photo-response (I-T) curves under 275 and 367 nm excitation wavelength is measured in Fig. 5(c)-5(d), and the photo-response is faster under 367 nm excitation light. Besides, we also tested the time-dependent photo-response curves under 532 and 700 nm excitation light in Fig. S6 and find good resolution, which demonstrates the potential for broadband detector.

To study the effect of phase transition $\boldsymbol{T}_1$~ 88 K on the performance of the MSM photodetector, and the photocurrent was measured at 78-98 K under the 375 nm excitation light (203 mW power from laser device) by Keithley 4200 in Fig. 5(e). The configuration of the MSM photodetector and dark current at low temperature is illustrated in Fig. S5 and Fig. S6. The photocurrent at 78 K is larger than that at other temperature, which aligns with the stronger PL absorption peak observed in Fig. 4(c). The ratio of $\boldsymbol{I_p/I_d}$ ($\boldsymbol{I}_{\mathrm{ratio}}$) at -5V above 83 K kept almost constant, and then increases ~ an order quickly below the phase transition at 78 K. Coincidently, the measured carrier lifetime is longer below 83 K in Fig. 5(f) and Fig. S6, which might correlate to the enhanced 3D EEI below 88 K[53]. With the increase of carrier lifetime, the photocurrent remains a lager value and the $\boldsymbol{I}_{\mathrm{ratio}}$ increases consequently, which demonstrates higher responsivity character after phase transition.

**Conclusion**

In conclusion, we have found two transitions in LSSO films at 88 K and 180 K, where the transition in 88 K could enhance carrier mobility and reduce carrier density at low temperature. Below $\boldsymbol{T}_1$, 3D EEI dominates, which causes different resistivity changing regularity and the transformation from negative MR to positive MR. photoluminescence (PL) spectra were tested to elucidate the phase transition and found a broad in-gap impurity absorption. Furthermore, we fabricated an MSM photodetector based on LSSO, which exhibits a wide response range from 275 nm to 700 nm, attributable to impurity levels, making it suitable for broadband spectrum detection. The observed increase in the photocurrent-to-dark current ratio after the T1 transition may be due to enhanced photocurrent and extended carrier lifetime. This study underscores the potential of leveraging impurity levels in perovskites for the advancement and design of broadband photodetectors.

**Acknowledgements**

This study was supported by the National Natural Science Foundation of China (Grants 12132020) to Yue Zheng, by National Natural Science Foundation of China (Grant 11904415 and 11972382) to Bangmin Zhang, by Guangdong Provincial Key Laboratory of Magnetoelectric Physics and Devices (No. 2022B1212010008) to Yue Zheng, by the Natural Science Foundation of Guangdong Province of China (Grant 2023A1515010882) and the Large Scientific Facility Open Subject of Songshan Lake, Dongguan, Guangdong (Grant KFKT2022B06) to Bangmin Zhang. The experiments reported were conducted on the Physical Research Platform in School of Physics, Sun Yat-sen University (PRPSP, SYSU).